# Fast GPGPU Data Rearrangement Kernels using CUDA

Michael Bader[1], Hans-Joachim Bungartz[1], Dheevatsa Mudigere[*,1,2], Srihari Narasimhan[2], Babu Narayanan[2]

[1]Chair for Scientific Computing, Department of Informatics
Technische Universität München,
Munich, Germany

[2]Computing & Decision Sciences lab
GE Global Research, JFWTC
Bangalore, India

*Abstract:* Many high performance computing algorithms are bandwidth limited, hence the need for optimal data rearrangement kernels as well as their easy integration into the rest of the application. In this work, we have built a CUDA library of fast kernels for a set of data rearrangement operations. In particular, we have built generic kernels for rearranging m dimensional data into n dimensions, including Permute, Reorder, Interlace/De-interlace, etc. We have also built kernels for generic Stencil computations on a two-dimensional data using templates and functors that allow application developers to rapidly build customized high performance kernels. All the kernels built achieve or surpass best-known performance in terms of bandwidth utilization.

## I. INTRODUCTION

In this paper, we address the problem of efficient data re-arrangement on the GPU, effectively utilizing the available bandwidth. Data re-arrangement forms a major bottleneck in majority of the basic computation kernels. It is very important to follow the right access pattern to get maximum memory bandwidth; especially given how costly accesses to device memory are, thus making it a non-trivial task. In case of many data intensive GPU applications there still exists a considerable gap, between the actual utilized bandwidth and the available bandwidth. Typically, large applications achieve upto 60% of the available bandwidth. An attempt has been made to bridge this gap, by developing speedy GPU kernels for basic data rearrangement operations. The target performance for each kernel is that it effectively utilizes upto 90% of the maximum bandwidth achievable. The maximum achievable bandwidth is taken as the maximum bandwidth attained for on device-to-device *memcpy* operations (77GB/s on the Tesla C1060). Furthermore these kernels are developed as generic implementations that allow for easy integration into existing applications.

There has been several previous works in literature directed towards improving performance of bandwidth-limited applications. Harrison et.al [1] have worked on Optimizing Data Movement Rates For Parallel Processing Applications On Graphics Processors using OpenGL, for the earlier generations of graphic cards. NVIDIA has explored data access dominated problems such as - a bandwidth efficient transpose kernel [2], optimal 3D finite difference kernel [3], efficient GPU implementation of a prefix scan operation [4] and sorting operation on GPU [5]. The open source *Thrust* [6] library provides C++ STL like functions for sorting, scan and other data manipulation primitives. Volkov et.al and the group from University of California at Berkeley, continue to actively work on optimal implementation of data intensive primitive linear algebra kernels on the GPU [7, 8].

## II. THE GPU AND CUDA

Graphic Processor Unit (GPU), has evolved into highly parallel, multithreaded, many-core coprocessors. The GPU is capable of sustaining tens of thousands of threads per application and offers tremendous computational power. The Tesla 10-series GPUs (Tesla C1060) contain 30 multi-processors; each multiprocessor contains 8 streaming processors (SM), in total 240 compute cores, offering a peak computing performance of upto one Teraflop. The Tesla C1060, has 4 GB of device memory with a theoretical max bandwidth of 102 GB/s.

NVIDIA's Compute unified device architecture (CUDA), provides a general purpose-programming model for GPUs. This has significantly boosted the use of GPUs for general purpose applications (GPGPU). The GPU kernels are driven from the CPU (host), offloading the compute intensive work onto the GPU. CUDA executes a kernel as a grid of thread-blocks. Threads are grouped into thread-blocks to allow for dividing the work among the available compute recourses. All threads in a thread-block can access any shared memory location assigned to that thread-block. Shared memory is limited fast memory available on each SM; its latency is two orders of magnitude lower than that of global memory. Global memory is far slower and data access operations to global memory usually limit performance of data intensive applications. To effectively utilize the bandwidth available it is required to follow to specific optimal data access patterns, such as – maintaining coalescence, avoiding partition camping and many others. Adhering to these access patterns is requirement for any performance kernel on the GPU. A more comprehensive and detailed description is provided in the CUDA programming guide [9]

### III. DATA REARRANGEMENT KERNELS

This refers to operations that involve rearranging data on the GPU device memory, such as – transpose, reorder, interlace/de-interlace etc. Data rearrangement operations inherently necessitate accessing data from different location in the memory. In case of the GPU, such a requirement presents a significant challenge to adhere to the optimal data access patterns.

The basic data-rearrangement operations that are being addressed as a part of this work are:

- **Basic read/write routines**: kernels to optimally read/write (access) data from the GPU global memory.
- **Data reordering routines**: reordering multi-dimensional data, given a specified sequence. *E.g: transposing a 2D matrix.*
- **Interlacing, de-interlacing of data**: operations that involve joining or splitting multi-dimensional data sets, according to a specified pattern. *E.g: splitting up the real and imaginary components for an array of complex numbers.*
- **Generic stencil computation kernel**: a generic kernel, for optimal stencil operations on two-dimensional data, given any specified stencil/convolution. *E.g: smoothing filter on a 2D image.*

### A. Basic read/write kernels

This is the primitive data re-arrangement operation. It is very necessary and important to clearly understand the concepts governing the performance for this operation. Data access (read/write) is a very basic operation and completely optimizing (solving) this would be a substantial effort in itself, and that would be out of the scope of the current work.

CUDA provides the *cudaMemcpy()* intrinsic function for the movement of data on the device. This intrinsic is used as the reference for comparison. Unlike the CUDA intrinsic memcpy function, this kernel allows for data transfer as per common access patterns. These kernels are templatized for the various access patterns, like accessing specified set of indices, access based on specified range, etc. One-dimensional CUDA blocks are used, with each block serviced by threads such that each thread handle four elements within a thread block *(vector computing model)*. The gridding and threading configuration is done automatically based on the data size. A comparison of the sequential data access pattern and the CUDA intrinsic memcpy, over a range of data sizes on the Tesla C1060 is given in Fig.1. The read kernel achieves a maximum bandwidth usage of 76 GB/s. The bandwidth usage of the read kernel is consistently greater than 95% of the bandwidth usage of the CUDA memcpy.

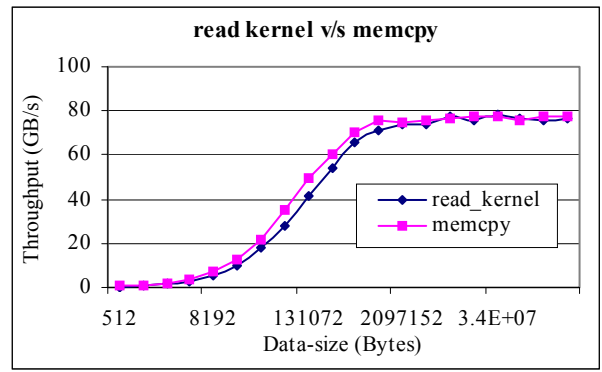

Fig. 1: Bandwidth utilization of the read kernel, Tesla C1060

### B. Data Re-ordering kernels

Generalized data-reordering operations involve rearranging N-dimensional data into M-dimensional data based on specified (any valid) reordering sequence. The basic storage order for N-dimensional data is given by - a vector called

'order'. This order vector contains a permutation of the numbers - 0, 1...N-1, with the fastest changing dimension coming first and followed by the successively slower changing dimensions. An N dimensional data set can be re-ordered in N-factorial possible ways. In the current work, row major linearized storage is used as the default for multi-dimensional data.

*3D Permute Kernel*: This kernel is to permute a given 3D dataset, there are six possible permutations of the ordering sequence - *[0 1 2], [0 2 1], [1 0 2], [1 2 0], [2 0 1]* and *[2 1 0]*. The 3D permutation is handled as a set of batched 2D data movement operations. Coalescing is maintained for global memory accesses. The 2D plane for the data movement operation is chosen based on the specified permutation order. Such that, it consists of the fastest changing dimensions of the input order and the desired (output) order. Block size of 32x32 elements is used, with 32x8 threads servicing each block. Every thread is responsible for four data elements *(vector computing model)*. A diagonalized ordering scheme for accessing the CUDA blocks is employed; this is to avoid the partition camping effects [10]. Table1 summarizes the performance results of the permute kernel on 128x256x512 data set, on the Tesla C1060. The maximum throughput for the different permutations is listed in table1.

It can be observed that the permute kernel attains upto 80-90% of the throughput of the *memcpy*. The variation in the bandwidth utilized for the different ordering sequences can be attributed to the different data access patterns used for each of these permutations.

| Permute order | Bandwidth (GB/s) Tesla C1060 |
|---|---|
| [0 1 2] *memcpy* | **77.82** |
| [0 2 1] | **62.55** |
| [1 0 2] | **63.17** |
| [1 2 0] | **57.38** |
| [2 0 1] | **59.63** |
| [2 1 0] | **58.42** |

Table. 1: 3D Permute kernel

*Reorder Kernel*: This kernel is for the more generic reorder operation. The 3D permute forms the building block for this kernel. The offset/striding approach is employed, to represent the different orders and convert between these orders in a generic way. The major issue in this case is to maintain coalescence during the data accesses. This poses a serious challenge, since the reorder operation inherently necessitates accessing data stored in non-contiguous memory locations. Similar to the permute kernels; two-dimensional blocks of size 32x32 are used, with 32x8 threads. The dimensions along which (2D) data are read in and written out are chosen such that coalescing is maintained during both these operations. This is accomplished by accessing data in a 2D plane defined by the fastest moving dimension of the original order (which is the dim - 0) and fastest moving dimension of the desired order. The above-discussed strategy holds good for *N-to-N* dimension reordering operations. But, with *N-to-M* reorder operations (*M < N*), maintaining coalescence during both read and write cannot be guaranteed in all cases. Particularly, when the desired order doesn't include the fastest changing dimension of the original order. In order to enhance performance of these kernels, the stride values are stored in the GPU-constant memory. This is advantageous since these stride values are recurrently accessed by all the threads. Additionally, the base index and the range, in case of the *N-to-M* reorder operation is also stored in the constant memory, as it is also accessed incessantly by all the threads.

The kernel takes in as arguments: the number of dimensions, an array of the sizes along each dimension, an array specifying the desired order and finally the input data. In case of the *N-to-M*, reorder operation an additional argument indicating the dimension of the output data is also passed to the kernel. Representative results from experiments conducted on the reorder kernel(s) on the Tesla C1060 are provided in the table2.

| Order | Data-size | Bandwidth (GB/s) Tesla C1060 |
|---|---|---|
| [1 0 2] | [256 256 256] (0.07 GB) | **76.00** |
| [1 0 2 3] | [256 256 256 1] (0.07 GB) | **75.41** |
| [3 2 0 1] | [256 256 1 256] (0.07 GB) | **56.24** |
| [3 0 2 1 4] | [256 16 1 256 16] (0.07 GB) | **43.40** |

Table. 2: Representative results for the generic reorder kernel

The performance of the kernel drops markedly for larger dimensions. A completely optimized and generic implementation of the kernel is an unrealizable goal, due to the limitations of the (fast) memory available on the device. But, the performance of the reorder kernel for lower dimension (<5) approaches nearly 85% of the performance of the *memcpy*.

## C. Interlace, De-Interlace Kernels

This is another popular data dominated operation wherein multiple (n) data-sets are interlaced together to form a single (interlaced) combined set of data or a single data-set is split into multiple (n) smaller, individual data-sets. As in the case of the other data rearrangement operations, this operation also inherently requires data stored in non-contiguous/distant parts of the memory to be accessed.

The shared memory is used as a buffer to hold the data and allow non-coalesced manipulation of data. Thus, ensuring the accesses to the global memory still remains coalesced. In case of the interlace kernel, each block reads the data from the global memory in a coalesced manner, into the shared memory. Here, it is re-arranged and split into individual arrays and these are written back to the global memory again in a coalesced manner. Similarly, also in case of the de-interlace kernel, the inverse operation is carried out in a similar way. The data is split into blocks of 8x8 and (n*64) threads are used to service these individual blocks, where *n* is the number of array being interlaced or the being split into eg: 2,3,8. Shared memory used by each kernel is equal to the sizes of (n*64) data elements. This is to store the elements during the intermediate steps. Representative results are provided in table3. It is very clear from the results that the interlace/de-interlace kernel on Tesla C1060 has achieved the target performance.

| Data-size (GB) | # arrays | Bandwidth (GB/s), Tesla C1060 | |
|---|---|---|---|
| | | Interlace Kernel | De-Interlace Kernel |
| 0.27 | 4 | **70.93** | **68.87** |
| 0.34 | 5 | **73.95** | **68.50** |
| 0.41 | 6 | **71.51** | **67.61** |
| 0.48 | 7 | **72.14** | **60.21** |
| 0.55 | 8 | **58.58** | **60.55** |
| 0.62 | 9 | **70.60** | **58.25** |

Table. 3: Representative results for Interlace/De-Interlace kernel

## D. Generic 2D Stencil Calculation Kernel

This is the final class of the data rearrangement operations. It is different from the earlier operations in the respect that this is not purely a data rearrangement operation and involves some amount of compute. 2D stencil computation, refers to computations involving (nearest) neighbors over a two dimensional grid. Each point in a two dimensional grid is updated with weighted contributions from a subset of its neighbors (spatial neighbors on the 2D grid) [3]. Such operations are very common and can be very widely found in many scientific computing applications, some examples are - PDE solvers, image filters etc. The dependency on neighboring data elements *(ghost zone/apron values)* reduces the extent of data parallelism - proving to be a possible bottleneck. The performance is commonly limited by the bandwidth, since majority of the operation consists of data-accesses.

This kernel provides a generic, optimal framework for stencil type computations. The actual calculations to be performed are dictated by the stencil. The actual required stencil is written as a Functor Object [11], with the single threaded version of the desired stencil function. The stencil kernel employs a 32x32 block with 32x8 threads, with each thread handling 4 elements within a block. Diagonalized ordering for the accessing the CUDA blocks is used to avoid partition camping effects. Further, similar to the previous kernels the shared memory is used as a user managed cache.

The stencil calculations of the elements at the border of the blocks require elements from neighboring blocs. Specifically designated threads handle this extra work of loading elements from neighboring blocks. For first order stencils - a thread block of 32x8 needs to load 34x34 elements. This introduces redundancy in the data being loaded by each block. So, there is an overlap of 32x4 elements between each of the blocks, barring the blocks at the boundary of the computational domain. This additional work results in warp divergence within the thread block, causing drop in performance of the kernel. Further, loading the additional ghost layers elements/apron-values is not coalesced, as they are beyond the scope of the block. This results in misaligned loads within the warp, resulting in drop in performance. But, the nature of the stencil operation makes these performance deterring operations essential.

In the current work, experiments are carried out with a (2D) finite difference stencil of different orders (I, II, III, IV). Detailed results from tests on the Tesla C1060 are provided in Fig.2. The 2D finite difference stencil is commonly used in case of discretized PDE solvers.

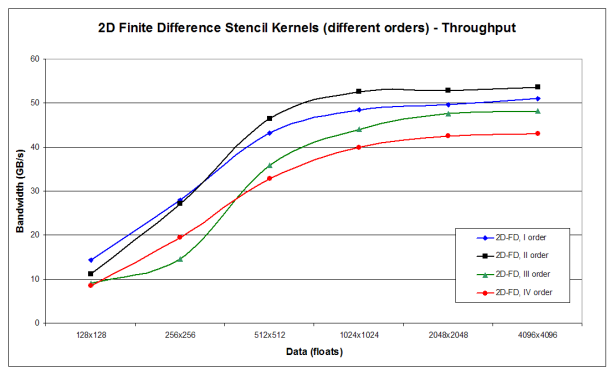

Fig. 2: Performance of 2D-FD stencil kernel, Tesla C1060

To avoid uncoalesced accesses - variants of the kernel that use texture memory have also been explored. Kernel variants with 1D, 2D texture memories and also a hybrid implementation that utilizes both the global and texture memories have been developed and compared (Table4). The coalesced data accesses are done through the global memory. Whereas the uncoalesced access to the apron values are handled though the texture memory. It can be seen from the table3, the use texture memory provides some improvement in performance but not very significant.

| Stencil kernel Variant | Bandwidth (GB/s) Tesla C1060 |
|---|---|
| Global memory | **51.07** |
| 1D Texture | **54.34** |
| Hybrid 1D Texture | **52.88** |
| 2D Texture | **47.22** |
| Hybrid 2D Texture | **53.91** |

Table. 4: Stencil Kernel variants with texture memory, I order 2D-FD stencil on 4096x4096 (float) data set

## IV. CONCLUSION

We have developed a library of optimal kernels for basic data rearrangement operations on the GPU. Each of these kernels has been hand-tuned and utilizes upto 85% of the CUDA intrinsic (device-device) *memcpy* function. Furthermore, these kernels have been built as generic kernels incorporating templates and functors. This generic structure allows for seamless inclusion of these kernels into existing applications. These kernels can be easily used as building blocks for larger data-intensive applications to improve the application's overall bandwidth utilization and hence the performance. To demonstrate this, we have implemented a 2D CFD flow solver on the GPU, which incorporates these data rearrangement kernels [12]. The performance of the CFD application has been greatly improved with an overall bandwidth utilization of 56 GB/s on the Tesla C1060. Furthermore, a 253x speedup over the serial CPU code (Intel Nehalem X5550, single core) and 13x speedup over the parallel CPU version (16 MPI processes on 8 cores of 2 Quad Nehalem X5550) has been observed.

Furthermore we intend to take this work forward, by developing optimal GPU implementations of additional data arrangement operations. One such immediate candidate is the generic multi-dimensional coordinate transformations (gridding operation). We intend to develop this into a more comprehensive and complete library of optimal GPU data rearrangement kernels.

ACKNOWLEDGEMENT

I would like to thank GE global research and JFWTC for the opportunity to carry out this work at their facility. This work is part of the master thesis submitted in partial fulfillment for the degree of Master of Science in Computational Science and Engineering at TUM, Munich Germany.